\documentclass{ecai}
 \usepackage{times}
 \usepackage{graphicx}
 \usepackage{latexsym}
 
 \usepackage[hyphens]{url}
 \usepackage{amsmath}
 \usepackage{amsfonts}       
 \usepackage{makecell}
 \usepackage{multirow}
 \usepackage{booktabs}       
 \usepackage{nicefrac}       
 \usepackage{microtype}      
 
 \newcommand{\bE}{\mathbb{E}}
 \newcommand{\bR}{\mathbb{R}}
 \newcommand{\cA}{\mathcal{A}}
 \newcommand{\cM}{\mathcal{M}}
 \newcommand{\cS}{\mathcal{S}}
 \newcommand{\cP}{\mathcal{P}}
 \newcommand{\cT}{\mathcal{T}}
 \newcommand{\cO}{\mathcal{O}}

 \newcommand{\vin}{V_i(\theta_1,\dots,\theta_N)}
 
 \newcommand{\vpn}{V^p(\theta_1,\dots,\theta_N; \theta_p)}
 \newcommand{\vinp}{V_i^p(\theta_1,\dots,\theta_N)}
 
 \newcommand{\vintot}{V_i^{\text{tot}}(\theta_1,\dots,\theta_N)}

 \newcommand{\vtn}{V(\theta_1,\dots,\theta_N)}
 \newcommand{\vtnnext}{V(\theta_1+\Delta\theta_1,\dots,\theta_N+\Delta\theta_N)}

 \newcommand{\sumT}{\sum_{t = 0}^{T}}

 \newcommand{\RTi}{\sumT\gamma^tr_i^t}
 \DeclareMathOperator*{\argmax}{arg\,max}
 
\begin{document}
 	
 	\title{Adaptive Mechanism Design: Learning to Promote Cooperation}
 	
 	\author{Tobias Baumann \and Thore Graepel \and John Shawe-Taylor \institute{University College London, UK, email: tobias.baumann.17@ucl.ac.uk} }
 	
 	\maketitle
 	\bibliographystyle{ecai}
 	
 	\begin{abstract}
		In the future, artificial learning agents are likely to become increasingly widespread in our society. They will interact with both other learning agents and humans in a variety of complex settings including social dilemmas. We consider the problem of how an external agent can promote cooperation between artificial learners by distributing additional rewards and punishments based on observing the learners' actions.
		We propose a rule for automatically learning how to create the right incentives by considering the players' anticipated parameter updates. Using this learning rule leads to cooperation with high social welfare in matrix games in which the agents would otherwise learn to defect with high probability. We show that the resulting cooperative outcome is stable in certain games even if the planning agent is turned off after a given number of episodes, while other games require ongoing intervention to maintain mutual cooperation. However, even in the latter case, the amount of necessary additional incentives decreases over time.
 	\end{abstract}
 	
 	\section{INTRODUCTION}
 	Social dilemmas highlight conflicts between individual and collective interests. Cooperation allows for better outcomes for all participants, but individual participants are tempted to increase their own payoff at the expense of others. Selfish incentives can therefore destabilize the socially desirable outcome of mutual cooperation and often lead to outcomes that make everyone worse off \cite{VanLange2013TheReview}.
 	
 	Cooperation often emerges due to direct reciprocity \cite{Trivers1971TheAltruism} or indirect reciprocity \cite{Nowak2005EvolutionReciprocity}. However, even if these mechanisms are not sufficient on their own, humans are often able to establish cooperation by changing the structure of the social dilemma. This is often referred to as \emph{mechanism design}. For instance, institutions	such as the police and the judicial system incentivize humans to cooperate in the social dilemma of peaceful coexistence, and have succeeded in dramatically reducing rates of violence \cite{Pinker2011TheNature}. 
 	
 	Studies of social dilemmas have traditionally focused on the context of human agents. However, in the future, artificial learning agents will likely be increasingly widespread in our society, and be employed in a variety of economically relevant tasks. In that case, they will interact both with other artificial agents and humans in complex and partially competitive settings.
 	
 	This raises the question of how we can ensure that artificial agents will learn to navigate the resulting social dilemmas productively and safely. Failing to learn cooperative policies would lead to socially inefficient or even disastrous outcomes. In particular, the escalation of conflicts between artificial agents (or between artificial agents and humans) may pose a serious security risk in safety-critical systems. The behaviour of artificial agents in cooperation problems is thus of both theoretical and practical importance. 
 	
 	
 	In this work, we will examine how mechanism design can promote beneficial outcomes in social dilemmas among artificial learners. We consider a setting with $N$ agents in a social dilemma and an additional \emph{planning agent} that can distribute (positive or negative) rewards to the players after observing their actions, and aims to guide the learners to a socially desirable outcome (as measured by the sum of rewards).
 	
 	We derive a learning rule that allows the planning agent to learn how to set the additional incentives by looking ahead at how the agents will update their policy parameter in the next learning step. 
 	We also extend the method to settings in which the planning agent does not know what internal parameters the other agents use and does not have direct access to the opponents' policy.
 	
 	We evaluate the learning rule on several different matrix game social dilemmas. The planning agent learns to successfully guide the learners to cooperation with high social welfare in all games, while they learn to defect in the absence of a planning agent. We show that the resulting cooperative outcome is stable in certain games even if the planning agent is turned off after a given number of episodes. In other games, cooperation is unstable without continued intervention. However, even in the latter case, we show that the amount of necessary additional rewards decreases over time.

 	\section{RELATED WORK}
 	The study of social dilemmas has a long tradition in game theory, theoretical social science, and biology. In particular, there is a substantial body of literature that fruitfully employs matrix games to study how stable mutual cooperation can emerge \cite{Axelrod1981TheCooperation}. Key mechanisms that can serve to stabilize the socially preferred outcome of mutual cooperation include direct reciprocity \cite{Trivers1971TheAltruism}, indirect reciprocity \cite{Nowak2005EvolutionReciprocity}, and norm enforcement \cite{Axelrod1986AnNorms}. \cite{bachrach2009cost} examine how cooperation can be stabilized via supplemental payments from an external party.
 	
 	
 	Our work is inspired by the field of \emph{mechanism design}, pioneered by \cite{Vickrey1961COUNTERSPECULATIONTENDERS}, which aims to design economic mechanisms and institutions to achieve certain goals, most notably social welfare or revenue maximization. \cite{Seabright1993ManagingDesign} studies how informal and formal incentives for cooperative behaviour can prevent a tragedy of the commons. \cite{monderer2004k} considers a setting in which an interested party can commit to non-negative monetary transfers, and studies the conditions under which desirable outcomes can be implemented with a given amount of payment. Mechanism design has also been studied in the context of computerized agents \cite{Varian1995EconomicAgents} and combined with machine learning techniques \cite{narasimhan2016automated}.
 	
 	We also draw on the rich literature on multi-agent reinforcement learning. It is beyond the scope of this work to review all relevant methods in multi-agent reinforcement learning, so we refer the reader to existing surveys on the subject \cite{Busoniu2008ALearning,Tuyls2012MultiagentProspects}. However, we note that most work in multi-agent reinforcement learning considers coordination or communication problems in the fully cooperative setting, where the agents share a common goal \cite{Omidshafiei2017DeepObservability,Foerster2016LearningLearning}.
 	
 	As an exception, \cite{Leibo2017Multi-agentDilemmas} study the learned behaviour of deep Q-networks in a fruit-gathering game and a Wolfpack hunting game that represent sequential social dilemmas. \cite{Tampuu2017MultiagentLearningb} successfully train agents to play Pong with either a fully cooperative, a fully competitive, or a mixed cooperative-competitive objective. \cite{Crandall2018CooperatingMachines} introduce a learning algorithm that uses novel mechanisms for generating and acting on signals to learn to cooperate with humans and with other machines in iterated matrix games. Finally, \cite{Lowe2017Multi-AgentEnvironments} propose a centralized actor-critic architecture that is applicable to both the fully cooperative as well as the mixed cooperative-competitive setting.
 	
 	However, these methods assume a given set of opponent policies as given in that they do not take into account how one's actions affect the parameter updates on other agents. In contrast, \cite{Foerster2017LearningAwareness} introduce Learning with Opponent-Learning Awareness (LOLA), an algorithm that explicitly attempts to shape the opponent's anticipated learning. The LOLA learning rule includes an additional term that reflects the effect of the agent's policy on the parameter update of the other agents and inspired the learning rule in this work. However, while LOLA leads to emergent cooperation in an iterated Prisoner's dilemma, the aim of LOLA agents is to shape the opponent's learning to their own advantage, which does not always promote cooperation.
 		
 	\section{BACKGROUND}
 		\subsection{Markov games}
 		We consider partially observable Markov games \cite{Littman1994MarkovLearning} as a multi-agent extension of Markov decision processes (MDPs). An $N$-player Markov game $\cM$ is defined by a set of states $\cS$, an observation function $O: \cS \times \{1,\dots,N\} \rightarrow \bR^d$ specifying each player's $d$-dimensional view, a set of actions $\cA_1, \dots, \cA_N$ for each player, a transition function $\cT: \cS \times \cA_1 \times \dots \times \cA_N \rightarrow \cP(\cS)$, where $\cP(\cS)$ denotes the set of probability distributions over $\cS$, and a reward function $r_i: \cS \times \cA_1 \times \dots \times \cA_N \rightarrow \bR$ for each player. To choose actions, each player uses a policy $\pi_i : \cO_i \rightarrow \cP(\cA_i)$, where $\cO_i = \{ o_i~|~s \in \cS, o_i = O(s,i)\}$ is the observation space of player $i$. Each player in a Markov game aims to maximize its discounted expected return $R_i = \RTi$, where $\gamma$ is a discount factor and $T$ is the time horizon.
 		
 		\subsection{Policy gradient methods} 
 		Policy gradient methods \cite{Sutton1998ReinforcementIntroduction} are a popular choice for a variety of reinforcement learning tasks. Suppose the policy $\pi_\theta$ of an agent is parametrized by $\theta$. Policy gradient methods aim to maximize the objective $J(\theta) = \bE_{s\sim p^{\pi_\theta},a\sim\pi_\theta}[R]$ by updating the agent's policy steps in the direction of $\nabla_\theta J(\theta)$.
 		
 		Using the policy gradient theorem \cite{SuttonPolicyApproximation}, we can write the gradient as follows:
 		\begin{equation}
 		\nabla_\theta J(\theta) = \bE_{s\sim p^{\pi_\theta},a\sim\pi_\theta}[\nabla_\theta \log \pi_\theta(a|s)\  Q^{\pi_\theta}(s,a)]
 		\end{equation}
 		where $p^{\pi_\theta}$ is the state distribution and $Q^{\pi_\theta}(s,a) = \bE[R| s_t = s, a_t = a]$.
 		
 		
 		\subsection{Matrix game social dilemmas}
 		\label{sec:mgsd}
 		A matrix game is the special case of two-player perfectly observable Markov games with $|\cS| = 1$, $T=1$ and $\cA_1 = \cA_2 = \{\text{C},\text{D}\}$. That is, two actions are available to each player, which we will interpret as cooperation and defection.
 		
 		\begin{table} 
 			\begin{center}
 			\caption{Payoff matrix of a symmetric 2-player matrix game. A cell of $X,Y$ represents a utility of $X$ to the row player and $Y$ to the column player.\label{fig:matrix_games}}
 			\begin{tabular}{c|c|c|c|}
 				& C & D \\
 				\hline
 				C & $R, R$ & $S, T$ \\
 				\hline
 				D & $T, S$ & $P, P$ \\
 				\hline
 			\end{tabular}
 			\end{center}
 		\end{table}
 	 		
 		Table \ref{fig:matrix_games} shows the generic payoff structure of a (symmetric) matrix game. Players can receive four possible rewards: $R$ (reward for mutual cooperation), $P$ (punishment for mutual defection), $T$ (temptation of defecting against a cooperator), and $S$ (sucker outcome of cooperating against a defector).
 		
 		A matrix game is considered a social dilemma if the following conditions hold \cite{Macy2002LearningDilemmas.}:
 		\begin{enumerate}
 			\item Mutual cooperation is preferable to mutual defection: $R > P$
 			
 			\item Mutual cooperation is preferable to being exploited: $R > S$
 			
 			\item Mutual cooperation is preferable to an equal probability of unilateral defection by either player: $R > \frac{T+S}{2}$
 			
 			\item The players have some reason to defect because exploiting a cooperator is preferable to mutual cooperation ($T > R$) or because mutual defection is preferable to being exploited ($P > S$). 
 			
 		\end{enumerate}
 		
 		The last condition reflects the mixed incentive structure of matrix game social dilemmas. We will refer to the motivation to exploit a cooperator (quantified by $T-R$) as \emph{greed} and to the motivation to avoid being exploited by a defector ($P-S$) as \emph{fear}. As shown in Table \ref{fig:matrix_games_examples}, we can use the presence or absence of greed and fear to categorize matrix game social dilemmas.
 		
 		\begin{table} 
 			\begin{center}
 			\caption{The three canonical examples of matrix game social dilemmas with different reasons to defect. In Chicken, agents may defect out of greed, but not out of fear. In Stag Hunt, agents can never get more than the reward of mutual cooperation by defecting, but they may still defect out of fear of a non-cooperative partner. In Prisoner's Dilemma (PD), agents are motivated by both greed and fear simultaneously. \label{fig:matrix_games_examples}}
 			\begin{tabular}{c|c|c|c|}
 				Chicken  & C & D \\
 				\hline
 				C & $3, 3$ & $1, 4$ \\
 				\hline
 				D & $4, 1$ & $0, 0$ \\
 				\hline
 			\end{tabular}
 			~~~~~~
 			\begin{tabular}{c|c|c|c|}
 				Stag Hunt & C & D \\
 				\hline
 				C & $4, 4$ & $0, 3$ \\
 				\hline
 				D & $3, 0$ & $1, 1$ \\
 				\hline
 			\end{tabular} 
 			~~~~~~
 			\begin{tabular}{c|c|c|c|}
 				PD & C & D \\
 				\hline
 				C & $3 ,3$ & $0, 4$ \\
 				\hline
 				D & $4, 0$ & $1, 1$ \\
 				\hline
 			\end{tabular}
 			\end{center}
 		\end{table}

 	\section{METHODS}
 		\subsection{Amended Markov game including the planning agent}
 		Suppose $N$ agents play a Markov game described by $\cS$, $\cA_1 \dots \cA_N$, $r_1,\dots,r_n$, $\cO$ and $\cT$. We introduce a \emph{planning agent} that can hand out additional rewards and punishments to the players and aims to use this to ensure the socially preferred outcome of mutual cooperation. 
 	
 		To do this, the Markov game can be amended as follows. We add another action set $\cA_p \subset \bR^N$ that represents which additional rewards and punishments are available to the planning agent. Based on its observation $\cO_p: \cS \times \{1,\dots,N\} \rightarrow \bR^d$ and the other player's actions $a_1,\dots,a_n$, the planning agent takes an action $a_p = (r_1^p,\dots, r_N^p) \in \cA_p \subset \bR^N$.\footnote{Technically, we could represent the dependence on the other player's actions by introducing an extra step after the regular step in which the planning agent chooses additional rewards and punishments. However, for simplicity, we will discard this and treat the player's actions and the planning action as a single step. Formally, we can justify this by letting the planning agent specify its action for every possible combination of player actions.} The new reward function of player $i$ is $r_i^{(tot)} = r_i + r_i^p$, i.e. the sum of the original reward and the additional reward, and we denote the corresponding value functions as $\vintot = \vin + \vinp$. Finally, the transition function $\cT$ formally receives $a_p$ as an additional argument, but does not depend on it ($\cT(s,a_1,\dots,a_N,a_p) = \cT(s,a_1,\dots,a_N)$).

 		\subsection{The learning problem}
 		Let $\theta_1,\dots,\theta_N$ and $\theta_p$ be parametrizations of the player's policies $\pi_1,\dots,\pi_N$ and the planning agent's policy $\pi_p$. 
 		
 		The planning agent aims to maximize the total social welfare $\vtn := \sum_{i=1}^N \vin$, which is a natural metric of how socially desirable an outcome is. Note that without restrictions on the set of possible additional rewards and punishments, i.e. $\cA_p = \bR^N$, the planning agent can always transform the game into a fully cooperative game by choosing $r_i^p = \sum_{j=1, j\neq i}^N r_j$. 
 		
 		However, it is difficult to learn how to set the right incentives using traditional reinforcement learning techniques. This is because $\vtn$ does not depend \emph{directly} on $\theta_p$. The planning agent's actions only affect $\vtn$ indirectly by changing the parameter updates of the learners. For this reason, it is vital to explicitly take into account how the other agents' learning changes in response to additional incentives.
 		
 		This can be achieved by considering the next learning step of each player (cf. \cite{Foerster2017LearningAwareness}). We assume that the learners update their parameters by simple gradient ascent:
 		\begin{equation}
 		\begin{aligned}
 		\Delta\theta_i &= \eta_i\nabla_i\vintot \\
 		&= \eta_i(\nabla_i\vin+\nabla_i\vinp)
 		\end{aligned}
 		\end{equation}
 		where $\eta_i$ is step size of player $i$ and $\nabla_i := \nabla_{\theta_i}$ is the gradient with respect to parameters $\theta_i$.
 		
 		Instead of optimizing $\vtn$, the planning agent looks ahead one step and maximizes $\vtnnext$. Assuming that the parameter updates $\Delta\theta_i$ are small, a first-order Taylor expansion yields
 		\begin{equation}
 		\begin{aligned}
 		&\vtnnext \approx \\
 		&\approx \vtn + \sum_{i=1}^N (\Delta\theta_i)^T \nabla_i\vtn
 		\end{aligned}
 		\end{equation}
 		We use a simple rule of the form $\Delta\theta_p = \eta_p \nabla_p\vtnnext$ to update the planning agent's policy, where $\eta_p$ is the learning step size of the planning agent. Exploiting the fact that $\vtn$ does not depend directly on $\theta_p$, i.e. $\nabla_p\vtn = 0$, we can  calculate the gradient:
 		\begin{equation}
 		\begin{aligned}
 		\nabla_p&\vtnnext \approx \\
 		&\approx \sum_{i=1}^N \nabla_p(\Delta\theta_i)^T \nabla_i\vtn \\
 		&= \sum_{i=1}^N \eta_i(\nabla_p\nabla_i\vintot)^T \nabla_i\vtn \\ 
 		&= \sum_{i=1}^N \eta_i(\nabla_p\nabla_i\vinp)^T \nabla_i\vtn \\ 
 		\end{aligned}
 		\label{eq:gradp_differentiate_through}
 		\end{equation}
 		since $\nabla_i\vin$ does not depend on $\theta_p$ either.

 		\subsection{Policy gradient approximation} 
 		If the planning agent does not have access to the exact gradients of $\vinp$ and $\vtn$, we use policy gradients as an approximation. Let $\tau = (s_0, \mathbf{a^{0}},a_p^0, \mathbf{r^{0}} \dots, s_T, \mathbf{a^{T}}, a_p^T, \mathbf{r^{T}})$ be a state-action trajectory of horizon $T+1$, where $\mathbf{a}^{t} = (a_1^t,\dots,a_N^t)$, $\mathbf{r}^{t} = (r_1^t,\dots,r_N^t)$, and $a_p^t = = (r_{1,p}^t,\dots,r_{N,p}^t)$ are the actions taken and rewards received in time step $t$. Then, the episodic return $R_i^0(\tau) = \RTi$ and $R_{i,p}^0(\tau) = \sum_{t=0}^T \gamma^t r_{i,p}^t$ approximate $\vin$ and $\vinp$, respectively. Similarly, $R^0(\tau) = \sum_{i=0}^N R_i^0(\tau)$ approximates the social welfare $\vtn$.
 		
 		We can now calculate the gradients using the policy gradient theorem:
 		\begin{equation}
 		\begin{aligned}
 		\nabla_i\vin &\approx \nabla_i\bE[R_i^0(\tau)] 	\\
 		&= \bE[\nabla_i\log \pi_i(\tau) R_i^0(\tau)] 
 		\end{aligned}
 		\end{equation}
 		The other gradients $\nabla_i\vtn$ and $\nabla_p\nabla_i\vinp$ can be approximated in the same way. This yields the following rule for the parameter update of the planning agent:
 		\begin{equation}
 		\begin{aligned}
 		\Delta\theta_p = \eta_p \sum_{i=1}^N \eta_i&\left(\bE\left[\nabla_p\log \pi_p(\tau) \nabla_i\log \pi_i(\tau) R_{i,p}^0(\tau)\right]\right)^T \\ 
 		\cdot &\bE\left[\nabla_i\log \pi_i(\tau) R^0(\tau)\right]
 		\end{aligned}
 		\label{eq:update_pg}
 		\end{equation}

 		\subsection{Opponent modeling}
 		Equations \ref{eq:gradp_differentiate_through} and \ref{eq:update_pg} assume that the planning agent has access to each agent's internal policy parameters and gradients. This is a restrictive assumption. In particular, agents may have an incentive to conceal their inner workings in adversarial settings. However, if the assumption is not fulfilled, we can instead model the opponents' policies using parameter vectors $\hat{\theta}_1,\dots,\hat{\theta}_N$ and infer the value of these parameters from the player's actions \cite{Ross2010ALearning}. A simple approach is to use a maximum likelihood estimate based on the observed trajectory:
 		\begin{equation}
 		\label{eq:opp_modeling}
 		\hat{\theta_i} = \argmax_{\theta_i^{'}} \sum_{t=0}^T \log \pi_{\theta_i^{'}}(a_t^i|s_t).
 		\end{equation}
 		Given this, we can substitute $\hat{\theta}_i$ for $\theta_i$ in equation \ref{eq:gradp_differentiate_through}. 
 		
 		
 		\subsection{Cost of additional rewards}
 		
 		In real-world examples, it may be costly to distribute additional rewards or punishment. We can model this cost by changing the planning agent's objective to $\vtnnext - \alpha ||\vpn||_2$, where $\alpha$ is a cost parameter and $V^p = (V_1^p,\dots,V_N^p)$. The modified update rule is (using equation \ref{eq:gradp_differentiate_through})
 		\begin{equation}
 		\label{cost_eq}
 		\small
 		\Delta\theta_p \!= \eta_p \!\!\left(\!\!\!\!
 		\begin{array}{r}
 		\displaystyle\sum_{i=1}^N \eta_i(\nabla_p\nabla_i\vinp)^T \nabla_i\vtn \\
 		- \alpha\nabla_p||\vpn||_2 
 		\end{array}
 		\!\!\!\right)
 		\end{equation}

 	\section{EXPERIMENTAL SETUP}
 	In our experiments, we consider $N = 2$ learning agents playing a matrix game social dilemma (MGSD) as outlined in section \ref{sec:mgsd}. The learners are simple agents with a single policy parameter $\theta$ that controls the probability of cooperation and defection: $P(C) = \frac{\exp(\theta)}{1+\exp(\theta)}$, $P(D)=\frac{1}{1+\exp(\theta)}$. The agents use a centralized critic \cite{Lowe2017Multi-AgentEnvironments} to learn their value function. 
 	
 	The agents play 4000 episodes of a matrix game social dilemma. We fix the payoffs $R=3$ and
 	$P=1$, which allows us to describe the game using the level of greed and fear. We will consider
 	three canonical matrix game social dilemmas as shown in Table \ref{fear_and_greed_table}.
 	
 	\begin{table} 
		\begin{center}
			\caption{Levels of fear and greed and resulting temptation $(T)$ and sucker $(S)$ payoffs in three matrix games. Note that the level of greed in Chicken has to be smaller than 1 because it is otherwise not a social dilemma ($R > \frac{T+S}{2}$ is not fulfilled). \label{fear_and_greed_table}}
			\begin{tabular}{c|c|c|c|c|c|}
				Game  & Greed & Fear & $T$ & $S$ \\
				\hline
				Prisoner's Dilemma & 1 & 1 & 4 & 0\\
				\hline
				Chicken & 0.5 & -1 & 3.5 & 2\\
				\hline
				Stag Hunt & -1 & 1 & 2 & 0\\
				\hline
			\end{tabular}
		\end{center}
 	\end{table}
 
 	The planning agent's policy is parametrized by a single layer neural network. We limit the maximum amount of additional rewards or punishments (i.e. we restrict $\cA_p$ to vectors that satisfy $\max_{i=1}^N |r_i^p| \leq c$ for a given constant $c$). 
 	Unless specified otherwise, we use a step size of 0.01 for both the planning agent and the learners, use cost regularisation (Equation \ref{cost_eq}) with a cost parameter of 0.0002, set the maximum reward to 3, and use the exact value function. In some experiments, we also require that the planning agent can only redistribute rewards, but cannot
 	change the total sum of rewards (i.e. $\cA_p$ is restricted to vectors that satisfy $\sum_{i=1}^N r_i^p = 0$). We refer to this as the \emph{revenue-neutral} setting.
 	
 	
 	
 		
 	\section{RESULTS}
 	In this section, we summarize the experimental results.\footnote{Source code available at \url{https://github.com/tobiasbaumann1/Adaptive_Mechanism_Design}} 
 	We aim to answer the following questions:
 	\begin{itemize}
 		\item Does the introduction of the planning agent succeed in promoting significantly higher levels of cooperation?
 		\item What qualitative conclusions can be drawn about the amount of additional incentives needed to learn and maintain cooperation?
 		\item In which cases is it possible to achieve cooperation even when the planning agent is only active for a limited timespan?
 		\item How does a restriction to revenue-neutrality affect the effectiveness of mechanism design?
 	\end{itemize}
 
 	\begin{figure*}[ht]
 		$$\begin{array}{cc}
 		\includegraphics[height=2in]{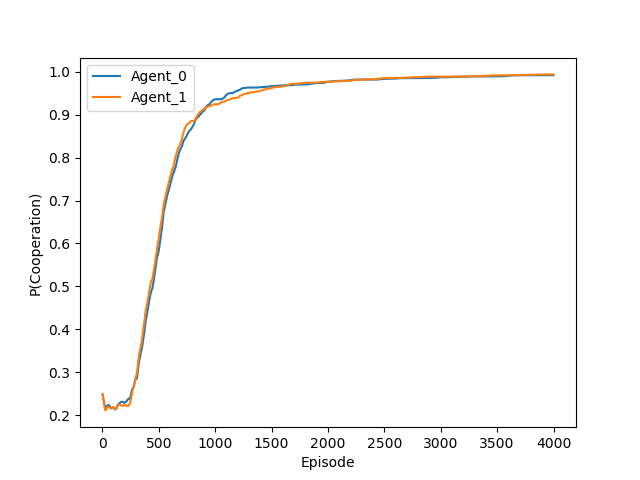} &
 		\includegraphics[height=2in]{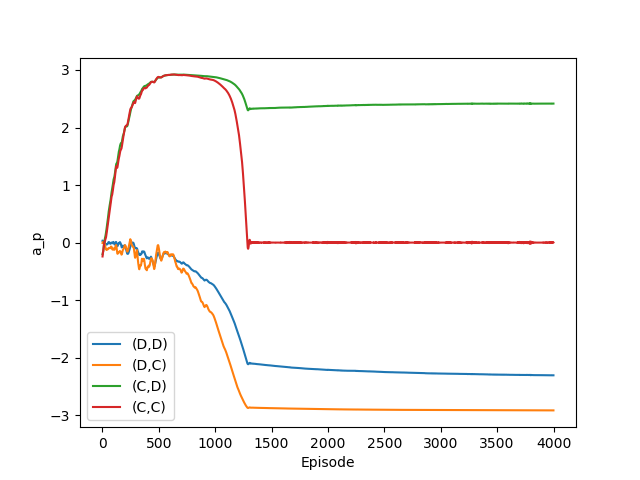} \\
 		\text{(a) Probability of cooperation}\label{fig:action_probs} & \text{(b) Additional rewards for player 1}\label{fig:planning_rewards} \\
 		\includegraphics[height=2in]{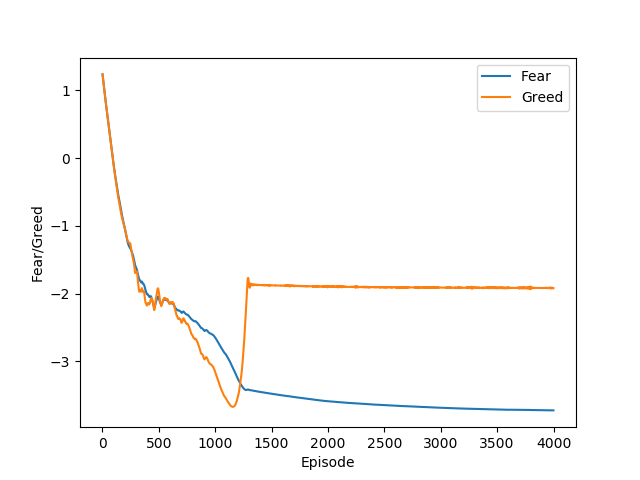} &
 		\includegraphics[height=2in]{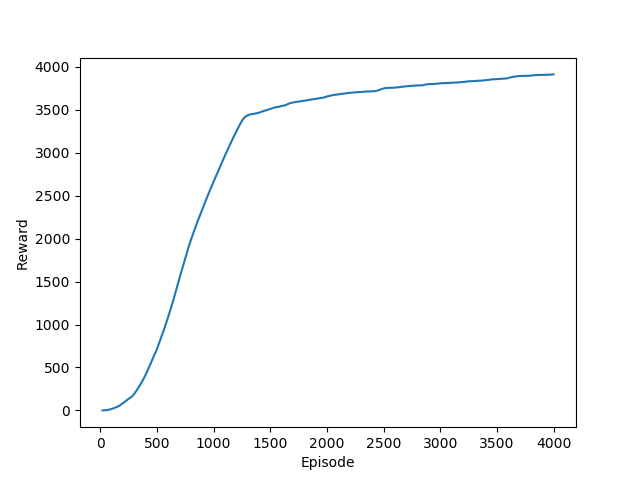} \\
 		\text{(c) Fear and greed in the modified game}\label{fig:fear_and_greed} & \text{(d) Cumulative additional rewards}	\label{fig:cum_planning_rewards}
 		\end{array}$$
 		\begin{center}
 			\caption{Mechanism design over 4000 episodes of a Prisoner's Dilemma. The initial probability of cooperation is 0.25 for each player. Shown is (a) the probability of cooperation over time, (b) the additional reward for the first player in each of the four possible outcomes, (c) the resulting levels of fear and greed including additional rewards, and (d) the cumulative amount of distributed rewards.}
 		\end{center}
 	\end{figure*}
 
 	Figure 1a illustrates that the players learn to cooperate with high probability if the planning agent is present, resulting in the socially preferred outcome of stable mutual cooperation. Thus the planning agent successfully learns how to distribute additional rewards to guide the players to a better outcome. 
 	
 	Figure 1b shows how the planning agent rewards or punishes the player conditional on each of the four possible outcomes. At first, the planning agent learns to reward cooperation, which creates a sufficient incentive to cause the players to learn to cooperate. In Figure 1c we show how this changes the level of fear and greed in the modified game. The levels of  greed and fear soon drop below zero, which means that the modified game is no longer a social dilemma. 
 	
 	Note that rewarding cooperation is less costly than punishing defection if (and only if) cooperation is the less common action. After the player learns to cooperate with high probability, the planning agent learns that it is now less costly to punish defection and consequently stops handing out additional rewards in the case of mutual cooperation outcome. As shown in Figure 1d, the amount of necessary additional rewards converges to 0 over time as defection becomes increasingly rare.
 	
 	Table \ref{comparison_table_1} summarizes the results of all three canonical social dilemmas. Without adaptive mechanism design, the learners fail to achieve mutual cooperation in all cases. By contrast, if the planning agent is turned on, the learners learn to cooperate with high probability, resulting in a significantly higher level of social welfare. 
 	\begin{table} 
 		\begin{center}
 			\caption{Comparison of the resulting levels of cooperation after 4000 episodes, a) without mechanism design, b) with mechanism design, and c) when turning off the planning agent after 4000 episodes and running another 4000 episodes. Each cell shows the mean and standard	deviation of ten training runs. $P(C,C)$ is the probability of mutual cooperation at the end of training and $V$ is the expected social welfare that results from the players' final action probabilities. The initial probability of cooperation is 0.25 for each player. \label{comparison_table_1}}
 			\begin{tabular}{cc|c|c|c|}
 				& & \makecell{Prisoner's \\ Dilemma} & Chicken & Stag Hunt \\ \hline
 				& Greed & 1 & 0.5 & -1 \\ \hline
 				& Fear & 1 & -1 & 1 \\ \hline \hline
 				\multirow{2}{1cm}{\centering No mech. design} & $P(C,C)$ & \makecell{0.004\% \\ $\pm$0.001\%} & \makecell{3.7\% \\ $\pm$1.3\%} & \makecell{0.004\% \\ $\pm$0.002\%} \\ \cline{2-5}
 				& $V$ & \makecell{2.024 \\ $\pm$0.003} & \makecell{5.44 \\ $\pm$0.01} & \makecell{2.00 \\ $\pm$0.00} \\ \hline \hline
 				\multirow{2}{1cm}{\centering With mech. design} & $P(C,C)$ & \makecell{98.7\% \\ $\pm$0.1\%} & \makecell{99.0\% \\ $\pm$0.1\%} & \makecell{99.1\% \\ $\pm$0.1\%} \\ \cline{2-5}
 				& $V$ & \makecell{5.975 \\ $\pm$0.002} & \makecell{5.995 \\ $\pm$0.001} & \makecell{5.964 \\ $\pm$0.005} \\ \hline \hline
 				\multirow{2}{1cm}{\makecell{Turning \\ off}} & $P(C,C)$ & \makecell{0.48\% \\ $\pm$0.4\%} & \makecell{53.8\% \\ $\pm$29.4\%} & \makecell{99.6\% \\ $\pm$0.0\%} \\ \cline{2-5}
 				& $V$ & \makecell{2.60 \\ $\pm$0.69} & \makecell{5.728 \\ $\pm$0.174} & \makecell{5.986 \\ $\pm$0.002} \\ \hline
 			\end{tabular}
 		\end{center}
 	\end{table}
 	
 	The three games differ, however, in whether the cooperative outcome obtained through mechanism design is stable even when the planning agent is turned off. Without additional incentives, mutual cooperation is not a Nash equilibrium in the Prisoner's Dilemma and in Chicken \cite{Fudenberg_Game_Theory}, which is why one or both players learn to defect again after the planning agent is turned off. These games thus require continued (but only occasional) intervention to maintain cooperation. By contrast, mutual cooperation is a stable equilibrium in Stag Hunt \cite{Fudenberg_Game_Theory}. As shown in Table \ref{comparison_table_1}, this means that long-term cooperation in Stag Hunt can be achieved even if the planning agent is only active over a limited timespan (and thus at limited cost). 
 	
 	\begin{table} 
		\begin{center}
			 \caption{Resulting levels of cooperation and average additional rewards (AAR) per round for different variants of the learning rule. The variants differ in whether they use the exact value function (Equation \ref{eq:gradp_differentiate_through}) or an estimate (Equation \ref{eq:update_pg}) and in whether the setting is revenue-neutral or unrestricted. \label{comparison_table_2}}
			\begin{tabular}{c|c|c|c|}
				& \makecell{Prisoner's \\ Dilemma} & Chicken & Stag Hunt \\ \hline
				Greed & 1 & 0.5 & -1 \\ \hline
				Fear & 1 & -1 & 1 \\ \hline \hline
				\multicolumn{4}{c|}{Exact $V$} \\ \hline
				$P(C,C)$ & \makecell{98.7\% \\ $\pm$0.1\%} & \makecell{99.0\% \\ $\pm$0.1\%} & \makecell{99.1\% \\ $\pm$0.1\%} \\ \hline
				AAR & \makecell{0.77 \\ $\pm$0.21} & \makecell{0.41 \\ $\pm$0.02} & \makecell{0.45 \\ $\pm$0.02} \\ \hline \hline
				\multicolumn{4}{c|}{\makecell{Exact $V$ \\ Revenue-neutral}} \\ \hline
				$P(C,C)$ & \makecell{91.4\% \\ $\pm$1.0\%} & \makecell{98.9\% \\ $\pm$0.1\%} & \makecell{69.2\% \\ $\pm$45.3\%} \\ \hline
				AAR & \makecell{0.61 \\ $\pm$0.04} & \makecell{0.31 \\ $\pm$0.02} & \makecell{0.19 \\ $\pm$0.11} \\ \hline \hline
				\multicolumn{4}{c|}{Estimated $V$} \\ \hline
				$P(C,C)$ & \makecell{61.3\% \\ $\pm$20.0\%} & \makecell{52.2\% \\ $\pm$18.6\%} & \makecell{96.0\% \\ $\pm$1.2\%} \\ \hline
				AAR & \makecell{3.31 \\ $\pm$0.63} & \makecell{2.65 \\ $\pm$0.31} & \makecell{4.89 \\ $\pm$0.39} \\ \hline
			\end{tabular}
		\end{center}
 	\end{table}
 	
 	Table \ref{comparison_table_2} compares the performance of different variants of the learning rule.	Interestingly, restricting the possible planning actions to redistribution leads to lower probabilities of cooperation in Prisoner's Dilemma and Stag Hunt, but not in Chicken. We hypothesize	that this is because in Chicken, mutual defection is not in the individual interest of the players anyway. This means  that the main task for the planning agent is to prevent (C,D) or (D,C) outcomes,
 	which can be easily achieved by redistribution. By contrast, these outcomes are fairly unattractive (in terms of individual interests) in Stag Hunt, so the most effective intervention is to make (D,D) less attractive and (C,C) more attractive, which is not feasible by pure redistribution. Consequently, mechanism design by redistribution works best in Chicken and worst in Stag Hunt.
 	 	
 	Using an estimate of the value function leads to inferior performance on all three games, both in
 	terms of the resulting probability of mutual cooperation and with respect to the amount of distributed additional results. However, the effect is by far least pronounced in Stag Hunt. This may be because mutual cooperation is an equilibrium in Stag Hunt, which means that a beneficial outcome can more easily arise even if the incentive structure created by the planning agent is imperfect.

 	Finally, we note that the presented approach is also applicable to settings with more than two players.\footnote{Source code available in a separate repository at \url{https://github.com/tobiasbaumann1/Mechanism_Design_Multi-Player}} We consider a multi-player Prisoner's Dilemma with $N=10$ agents.\footnote{The payoffs are as follows: 3 if all players cooperate, 1 if all players defect, 4 if you are the only to defect, 0 if you are the only to cooperate. Payoffs of intermediate outcomes, where some fraction of players cooperate, are obtained by linear interpolation.} 
 	\begin{figure*}[ht]
 		$$\begin{array}{cc}
 		\includegraphics[height=2in]{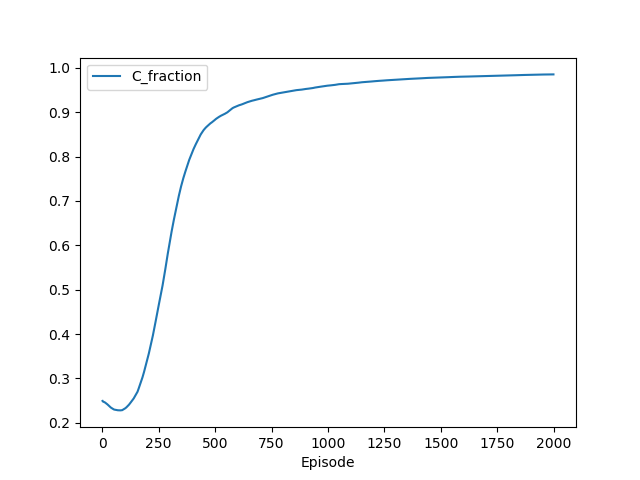} & \includegraphics[height=2in]{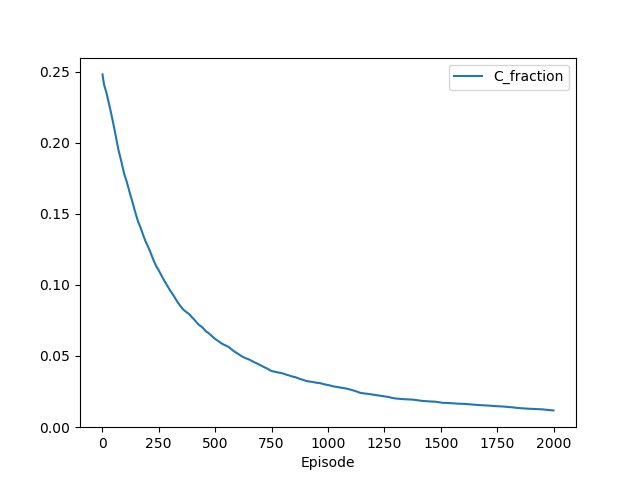} \\
 		\makecell{\text{(a) Average probability of cooperation with} \\ \text{mechanism design}} \label{fig:coop_probs_with_MD} & \makecell{\text{(b) Average probability of cooperation without} \\ \text{mechanism design}} \label{fig:coop_prob_without_MD}
 		\end{array}$$
 		\caption{Mechanism design in a multi-player Prisoner's Dilemma. The initial probability of cooperation is 0.25 for each player. Shown is the average probability of cooperation over time (a) in the presence of a planning agent, (b) without mechanism design.}
 	\end{figure*}
 	Figure 2a illustrates that, just as in the case of $N=2$, the players learn to cooperate with high probability if the planning agent is present. By contrast, without mechanism design, the players (unsurprisingly) converge to the socially undesirable outcome of mutual defection. This shows that the presented approach for learning how to distribute additional rewards scales easily to multi-agent social dilemmas.

 	\section{CONCLUSIONS AND FUTURE WORK}
 	We have presented a method for learning how to create the right incentives to ensure cooperation
 	between artificial learners. Empirically, we have shown that a planning agent that uses the proposed learning rule is able to successfully guide the learners to the socially preferred outcome of mutual cooperation in several different matrix game social dilemmas, while they learn to defect with high probability in the absence of a planning agent. The resulting cooperative outcome is stable in certain games even if the planning agent is turned off after a given number of episodes, while other games require continued (but increasingly rare) intervention to maintain cooperation. We also showed that restricting the planning agent to redistribution leads to worse performance in Stag Hunt, but not in Chicken.
 	
 	In the future, we would like to explore the limitations of adaptive mechanism design in more complex
 	environments, particularly in games with more than two players, without full observability of the
 	players' actions, and using opponent modeling (cf. Equation \ref{eq:opp_modeling}). Future work could also consider settings in which the planning agent aims to ensure
 	cooperation by altering the dynamics of the environment or the players' action set (e.g. by
 	introducing mechanisms that allow players to better punish defectors or reward cooperators).
 	
 	Finally, under the assumption that artificial learners will play vital roles in future society, it is worthwhile to develop policy recommendations that would facilitate mechanism design for these agents (and the humans they interact with), thus contributing to a cooperative outcome in potential social dilemmas. For instance, it would be helpful if the agents were set up in a way that makes their intentions as transparent as possible and allows for simple ways to distribute additional rewards and punishments without incurring large costs.

  	 		
	\bibliography{References}

\begin{thebibliography}{10}

\bibitem{Axelrod1986AnNorms}
Robert Axelrod, `{An Evolutionary Approach to Norms}', {\em American Political
  Science Review}, (1986).

\bibitem{Axelrod1981TheCooperation}
Robert Axelrod and William~D. Hamilton, `{The Evolution of Cooperation}', {\em
  Evolution}, (1981).

\bibitem{bachrach2009cost}
Yoram Bachrach, Edith Elkind, Reshef Meir, Dmitrii Pasechnik, Michael
  Zuckerman, J{\"o}rg Rothe, and Jeffrey~S Rosenschein, `The cost of stability
  in coalitional games', in {\em International Symposium on Algorithmic Game
  Theory}, pp. 122--134. Springer, (2009).

\bibitem{Busoniu2008ALearning}
Lucian Busoniu, Robert Babuska, and Bart De~Schutter, `{A Comprehensive Survey
  of Multiagent Reinforcement Learning}', {\em Systems, Man, and Cybernetics,
  Part C: Applications and Reviews}, (2008).

\bibitem{Crandall2018CooperatingMachines}
Jacob~W. Crandall, Mayada Oudah, Fatimah Ishowo-Oloko, Sherief Abdallah,
  Jean-Fran{\c{c}}ois Bonnefon, et~al., `Cooperating with machines', {\em
  Nature communications}, {\bf 9}(1),  233, (2018).

\bibitem{Foerster2016LearningLearning}
Jakob Foerster, Ioannis~Alexandros Assael, Nando de~Freitas, and Shimon
  Whiteson, `{Learning to Communicate with Deep Multi-Agent Reinforcement
  Learning}',  2137--2145, (2016).

\bibitem{Foerster2017LearningAwareness}
Jakob~N. Foerster, Richard~Y. Chen, Maruan Al-Shedivat, Shimon Whiteson, Pieter
  Abbeel, and Igor Mordatch, `{Learning with Opponent-Learning Awareness}',
  (2017).

\bibitem{Fudenberg_Game_Theory}
Drew Fudenberg and Jean Tirole, {\em {Game Theory}}, MIT Press, Cambridge, MA,
  1991.

\bibitem{Leibo2017Multi-agentDilemmas}
Joel~Z. Leibo, Vinicius Zambaldi, Marc Lanctot, Janusz Marecki, and Thore
  Graepel, `{Multi-agent Reinforcement Learning in Sequential Social
  Dilemmas}', {\em Proceedings of the 16th Conference on Autonomous Agents and
  MultiAgent Systems}, (2017).

\bibitem{Littman1994MarkovLearning}
Michael~L. Littman, `{Markov games as a framework for multi-agent reinforcement
  learning}', in {\em Machine Learning Proceedings 1994}, (1994).

\bibitem{Lowe2017Multi-AgentEnvironments}
Ryan Lowe, Yi~Wu, Aviv Tamar, Jean Harb, Pieter Abbeel, and Igor Mordatch,
  `Multi-agent actor-critic for mixed cooperative-competitive environments', in
  {\em Advances in Neural Information Processing Systems}, pp. 6382--6393,
  (2017).

\bibitem{Macy2002LearningDilemmas.}
Michael~W. Macy and Andreas Flache, `{Learning Dynamics in Social Dilemmas.}',
  {\em Proceedings of the National Academy of Sciences of the United States of
  America}, (2002).

\bibitem{monderer2004k}
Dov Monderer and Moshe Tennenholtz, `k-implementation', {\em Journal of
  Artificial Intelligence Research}, {\bf 21},  37--62, (2004).

\bibitem{narasimhan2016automated}
Harikrishna Narasimhan, Shivani~Brinda Agarwal, and David~C Parkes, `Automated
  mechanism design without money via machine learning', (2016).

\bibitem{Nowak2005EvolutionReciprocity}
Martin~A. Nowak and Karl Sigmund.
\newblock {Evolution of Indirect Reciprocity}, 2005.

\bibitem{Omidshafiei2017DeepObservability}
Shayegan Omidshafiei, Jason Pazis, Christopher Amato, Jonathan~P. How, and John
  Vian, `{Deep Decentralized Multi-task Multi-Agent Reinforcement Learning
  under Partial Observability}', (2017).

\bibitem{Pinker2011TheNature}
Steven Pinker, `{The Better Angels of Our Nature}', (2011).

\bibitem{Ross2010ALearning}
Stephane Ross, Geoffrey~J. Gordon, and J.~Andrew Bagnell, `{A Reduction of
  Imitation Learning and Structured Prediction to No-Regret Online Learning}',
  (2010).

\bibitem{Seabright1993ManagingDesign}
Paul Seabright, `{Managing Local Commons: Theoretical Issues in Incentive
  Design}', {\em Journal of Economic Perspectives}, {\bf 7}(4),  113--134,
  (1993).

\bibitem{SuttonPolicyApproximation}
Richard~S. Sutton, David~A. McAllester, Satinder~P. Singh, and Yishay Mansour,
  `Policy gradient methods for reinforcement learning with function
  approximation',  1057--1063, (2000).

\bibitem{Sutton1998ReinforcementIntroduction}
RS~Sutton and AG~Barto, {\em {Reinforcement learning: An introduction}}, 1998.

\bibitem{Tampuu2017MultiagentLearningb}
Ardi Tampuu, Tambet Matiisen, Dorian Kodelja, Ilya Kuzovkin, Kristjan Korjus,
  Juhan Aru, Jaan Aru, and Raul Vicente, `{Multiagent cooperation and
  competition with deep reinforcement learning}', {\em PLoS ONE}, (2017).

\bibitem{Trivers1971TheAltruism}
Robert~L. Trivers, `{The Evolution of Reciprocal Altruism}', {\em The Quarterly
  Review of Biology}, (1971).

\bibitem{Tuyls2012MultiagentProspects}
Karl Tuyls and Gerhard Weiss, `{Multiagent Learning: Basics, Challenges, and
  Prospects}', {\em AI Magazine}, (2012).

\bibitem{VanLange2013TheReview}
Paul A.~M. Van~Lange, Jeff Joireman, Craig~D. Parks, and Eric Van~Dijk, `{The
  Psychology of Social Dilemmas: A Review}', {\em Organizational Behavior and
  Human Decision Processes}, {\bf 120}(2),  125--141, (2013).

\bibitem{Varian1995EconomicAgents}
Hal~R. Varian, `Economic mechanism design for computerized agents.', in {\em
  USENIX workshop on Electronic Commerce}, pp. 13--21, (1995).

\bibitem{Vickrey1961COUNTERSPECULATIONTENDERS}
William Vickrey, `{Counterspeculation, Auctions, and Competitive Sealed
  Tenders}', {\em The Journal of Finance}, {\bf 16}(1),  8--37, (1961).

\end{thebibliography}
 	\end{document}